\documentstyle[aps,psfig]{revtex}

\begin{document}
\twocolumn

\title{Two-neutron removal reactions of $^6$He treated as a three-body halo} 

\author{\sc E.~Garrido} \address{Instituto de
Estructura de la Materia, CSIC, Serrano 123, E-28006 Madrid, Spain }
\author{{\sc D.V.~Fedorov} and {\sc A.S.~Jensen}}

\address{Institute of Physics
and Astronomy, Aarhus University, DK-8000 Aarhus C, Denmark}

\maketitle
\begin{abstract}
{\em Abstract.} We formulate a method to compute breakup processes of
three-body halo systems reacting with a target. The reaction between one
of the particles and the target as well as the corresponding pairwise
final state interactions are carefully treated. Both absorption and
diffraction processes are included. Detailed differential and total
cross sections are calculated for two-neutron (2n) removal from $^6$He
by fragmentation on $^{12}$C.  Effects due to the core shadowing are
taken into account. Good agreement with available experimental data
is obtained.\\
25.60.-t -- Reactions induced by unstable nuclei\\
25.60.Gc -- Break--up and momentum distributions\\
21.45.+v -- Few--body systems\\
\end{abstract}

{\it Introduction.} Breakup reactions of three-body systems are
difficult to describe in general if the formulation is required to be
both practical and fairly accurate. However, due to the steady
improvement of radioactive beam facilities an increasing amount of
fragmentation data demand analyses where interpretation in terms of
few-body reactions appear to be an efficient framework
%\cite{han95,ann90,zin95,nil95,hum95,ale97}. Previous attempts are
 \cite{han95,zin95,nil95,hum95,ale97}. Previous attempts are
based on the eikonal approximation and the adiabatic assumption of
frozen intrinsic degrees of freedom during the collision
%\cite{suz94,zhu94,kor94,for95,zhu95,gar96,gar97,alk96,ber97}.  In most
 \cite{suz94,zhu94,kor94,for95,gar96,gar97,alk96,ber97}.  In most
of these approaches very simple three-body wave functions are used and
the final state interaction often needs better treatment.

The crucial ingredients in these processes are the final state
interaction \cite{zin95,kor94,bar93} and the interaction between one
of the three particles (the participant) and the target. The remaining
two particles are essentially spectators and reactions simultaneously
involving more than one particle are much less important. The large
spatial extension of halo nuclei guarantees that this model is a good
first approximation which could be improved in future more elaborate
investigations if needed. We shall here fully exploit these
simplifying assumptions to keep the approach practical enough to
compute many observables.  We shall describe the reaction between each
particle and the target by the optical model. The final state two-body
interactions are those used to compute the bound state wave function
of the three-body projectile.

{\it Theory.}  When a spatially extended three body halo hits a
relatively small target at high energy the probability that more than
one of the constituents interacts strongly with the target is
small. The cross section can then be written as a sum of three terms
where each term describes the contribution to the reaction caused by
the interaction between the target and the corresponding particle

\begin{equation}
d\sigma =\sum_{i=1}^{3}d\sigma ^{(i)}=\sum_{i=1}^{3}\frac{1}{v}\frac{2\pi }{
\hbar }|T^{(i)}|^{2}d\nu _{f}^{(i)},
\end{equation}
where $v$ is the relative projectile-target velocity, $T^{(i)}$ is the
transition matrix element and $d\nu _{f}^{(i)}$ is the density of
final states. Each of the three terms includes elastic and inelastic
scattering of the corresponding particle on the target. 
\begin{figure}
\psfig{file=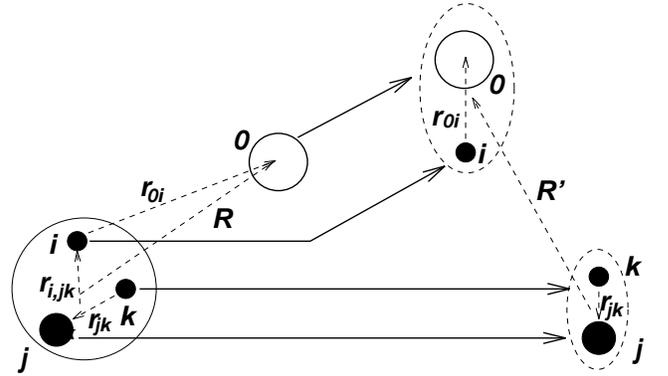,width=3.3in,angle=-90}
\vspace{0.5cm}
\caption{\footnotesize Sketch of the reaction and the coordinates
used. The target is labelled by 0 and $\{i,j,k\}$ label the particles
within the three-body projectile.  }
\end{figure}

Neglecting the Coulomb interaction and assuming that the target has
zero spin we write the transition matrix element for the elastic
contribution as (compare with the formulation for a weakly bound
projectile \cite{ban67})
\begin{eqnarray}
&&T^{(i)}= \langle \phi _{{\bf p}
_{0i}^{\prime } \Sigma^{\prime }_i}^{0i(+)} \phi _{{\bf p}
_{jk}^{\prime }s_{jk} \Sigma_{jk}}^{jk (+)} e^{i{\bf P}^{\prime }
{\bf R}^{\prime } }|V_{0i}|\Psi^{JM} 
e^{i{\bf P}{\bf R}} \rangle \nonumber\\
&&= \sum_{\Sigma_{i}} T^{(0i)}_{\Sigma_i \Sigma^{\prime}_i} 
M_{s_{jk} \Sigma_{jk} \Sigma_{i}}^{JM} \; , \\
&&  T^{(0i)}_{\Sigma_i \Sigma^{\prime}_i} =
\langle \phi _{{\bf p}
_{0i}^{\prime } \Sigma^{\prime}_i}^{0i (+)}|V_{0i}|
e^{i{\bf p}_{0i}{\bf r}_{0i}}
  \chi_{s_i \Sigma_{i}} \rangle \; , \\
&&M_{s_{jk} \Sigma_{jk} \Sigma_i}^{JM} = \langle
\phi _{{\bf p}_{jk}^{\prime }s_{jk} \Sigma_{jk}}^{jk(+)} e^{i
{\bf p}_{i,jk}{\bf r}_{i,jk}}\chi _{s_{i}\Sigma _{i}}|\Psi^{JM}\rangle \; ,
\end{eqnarray}
where $V_{0i}$ is the interaction between particle $i$ (the
participant) and the target, $\Psi^{JM}$ is the projectile internal
wave function, $\phi _{ {\bf p}_{0i}^{\prime }\Sigma
^{\prime}_i}^{0i(+)}$ and $\phi _{{\bf p}_{jk}^{\prime }s_{jk}
\Sigma_{jk}}^{jk(+)}$ are the distorted waves in the corresponding
two-body subsystems and $\chi _{s_{i}\Sigma^{\prime } _{i}}$, $\chi
_{s_{jk}\Sigma _{jk}}$ are the spin functions ($ {\bf s}_{jk}={\bf
s}_{j}+{\bf s}_{k}$). The coordinates ${\bf r}_{jk}, {\bf R}$ and
${\bf R}^{\prime}$ are defined in fig. 1 and the conjugate momenta are
denoted by the corresponding ${\bf p}$. Primes are used for the final
states.  The density of final states is
\begin{equation}
d\nu _{f}^{(i)} =   \delta (E_{0i}^{\prime }-E_{0i})
\frac{d^{3}{\bf p}_{0i}^{\prime }}{(2\pi
\hbar )^{3}}\frac{d^{3}{\bf p}_{jk}^{\prime }}{(2\pi \hbar )^{3}}\frac{d^{3}
{\bf P}^{\prime }}{(2\pi \hbar )^{3}} \; ,
\end{equation}
where $E_{0i}$ is the relative energy of particle $i$ and the target and
\begin{eqnarray}
&&{\bf r}_{0i}={\bf R-}\frac{m_{j}+m_{k}}{m_{i}+m_{j}+m_{k}}{\bf r}_{i,jk} 
, \\  
&&{\bf p}_{i,jk} ={\bf P}^{\prime }-\frac{m_{j}+m_{k}}{m_{i}+m_{j}+m_{k}}{\bf P}
, \\ 
&&{\bf p}_{0i} = {\bf P} - \frac{m_{0}}{m_{i}+m_{0}}{\bf P}^{\prime }.
\end{eqnarray}

The diffraction cross section for non-polarized projectile is then
obtained after summing over final and averaging over initial states
\begin{eqnarray} 
&&\frac{d^9\sigma _{el}^{(i)}({\bf P}^{\prime },{\bf 
p}_{jk}^{\prime },{\bf p}_{0i}^{\prime })}
{ d{\bf P}^{\prime } d{\bf p}_{jk}^{\prime } d{\bf p}_{0i}^{\prime } }
=\frac{d^3\sigma _{el}^{(0i)}({\bf p}_{0i}
  \rightarrow {\bf p}_{0i}^{\prime})}
 {d{\bf p}_{0i}^{\prime }}
  \frac{1}{2 J+1} \nonumber \\
&&\times\sum_{M s_{jk}\Sigma_{jk}\Sigma_i }
 |M_{s_{jk} \Sigma_{jk} \Sigma_i}^{JM}|^{2}  \label{eq5} \\
&&\frac{d^3\sigma _{el}^{(0i)}({\bf p}_{0i}
  \rightarrow {\bf p}_{0i}^{\prime})} {d{\bf p}_{0i}^{\prime }} 
=\frac{1}{v}\frac{2\pi }{\hbar } \frac{1}{2s_i+1}\nonumber\\
&&\times \sum_{\Sigma_i\Sigma^{\prime}_i} 
|T^{(0i)}_{\Sigma_i \Sigma^{\prime}_i}|^2
\delta (E_{0i}^{\prime }-E_{0i})\frac{d^{3}{\bf p}_{0i}^{\prime }}{(2\pi
\hbar )^{3}}  \; .
\end{eqnarray}

The absorption cross section where only two projectile particles
survive in the final state is simply obtained by replacing the elastic
$\sigma _{el}$ by $\sigma _{abs}$, i.e.
\begin{eqnarray} 
&&\frac{d^6\sigma _{abs}^{(i)}({\bf P}^{\prime },{\bf 
p}_{jk}^{\prime })}
{ d{\bf P}^{\prime } d{\bf p}_{jk}^{\prime } }
= \sigma _{abs}^{(0i)}(p_{0i})
  \frac{1}{2 J+1} \nonumber\\ 
&&\times\sum_{M s_{jk}\Sigma_{jk}\Sigma_i }
 |M_{s_{jk} \Sigma_{jk} \Sigma_i}^{JM}|^{2}
\label{eq6}
\end{eqnarray}
The total cross section arising from $V_{0i}$ is the sum of
eqs.(\ref{eq5}) and (\ref{eq6}).  The function $M_{s_{jk}
\Sigma_{jk},\Sigma_i}^{JM}$ is the transition matrix computed in
\cite{gar97}.

{\it Parameters.} We now consider the nucleus $^6$He (n+n+$\alpha$)
with the wave function obtained by solving the Faddeev equations in
coordinate space \cite{gar96} by using the potentials from \cite{cob97}.
The resulting three-body wave function has 88\% of p$^2$- and 12\% of
s$^2$-configurations. The binding energy is 0.95 MeV and the root mean
square radius is 2.45 fm.

For the neutron-target interactions we use non-relativistic
optical potentials obtained from relativistic potentials through a
reduction of the Dirac equation into a Schr\"{o}dinger-like equation
\cite{udi95}. In particular we focus
on a carbon target, and for the neutron--$^{12}$C interaction we use
the parametrization EDAI-C12 \cite{coo93} valid for a range of neutron
energies from 29 to 1040 MeV.  We include 35 partial waves in the
calculations. 

The detected 2n removal contribution from the interaction between
$\alpha$-particle and target is expected to be relatively small. First
the elastic scattering cross section for $\alpha$-particles on
$^{12}$C is about 35\% of the total cross section \cite{pen81},
which in turn is of the same order as the neutron $^{12}$C total cross
section. Second a substantial fraction of these 35\%, where the
$\alpha$-particle survives the reaction with the same energy, are in
fact elastically scattered $^6$He-particles and as such not
contributing to the 2n removal cross section. Third the very forward
angles, containing most of the elastic cross section, are
experimentally excluded. We shall therefore neglect this contribution
in the present letter.

The final state two-body continuum wave functions
$(\phi^{jk(+)},\phi^{0i(+)})$ are now calculated with the appropriate
boundary conditions and normalization \cite{gar97}. Then
eqs.(\ref{eq5}) and (\ref{eq6}) are integrated over the unobserved
quantities and different momentum distributions or differential cross
sections are obtained. However, the so-called core shadowing problem
remains \cite{esb96,han96,hen96}. Basically this means that the finite
sizes of the target and the $\alpha$-core prohibit some of the
neutron-target reactions namely those where the $\alpha$-particle also
interacts with the target.  In the eikonal approximation this effect
is accounted for by multiplying the halo wave function by a profile
function. A simpler approximation is the black disk model where the
profile functions are step functions \cite{bar93}, and the interior
part of the halo wave function is eliminated. This amounts to
eliminate the configuration which is not consistent with the reaction
in question.

In particular, in a process where the neutron is removed by the target
without destroying the $\alpha$-core, we must exclude the part of the
three-body wave function, where the removed neutron is too close (or
even inside) the core. We do this by assuming that $\Psi_i=0$ when the
distance between the interacting neutron and the center-of-mass of the
remaining two projectile particles is sufficiently small. This amounts
to assuming that the Jacobi coordinate $y=r_{n,\alpha
n}\sqrt{(m_\alpha+m_n)/(m_\alpha+2m_n)}$ is smaller than a certain cut
off value $y_c$, which presumably is better left as a parameter, but
clearly is related to the sizes of the reacting particles.  In fact,
the sum of the target and core radii is around 4 fm, and we expect
that for $r_{n,\alpha n}<4$ fm the neutron can not be removed without
disturbing (or breaking) the core.

{\it Observables.} In fig. 2 we compare the calculated and
experimental \cite{ale97} transverse $^5$He momentum distribution
after $^6$He fragmentation on $^{12}$C at 240 MeV/u. Without shadowing
a too broad momentum distribution is found, while the shadowing with
$y_c=6$ fm gives a too narrow distribution. For the estimated value of
$y_c=4$ fm, the central part of the distribution up to about 60 MeV/c
is rather well reproduced. The momentum tails receive contributions
from distances where the neutron is inside the core. The model is in
other words not applicable for these momenta. Thus $y_c=4$ fm seems to
be an appropriate value. This is consistent with the interpretation in
\cite{ale97,ber97}.

\begin{figure}
\psfig{file=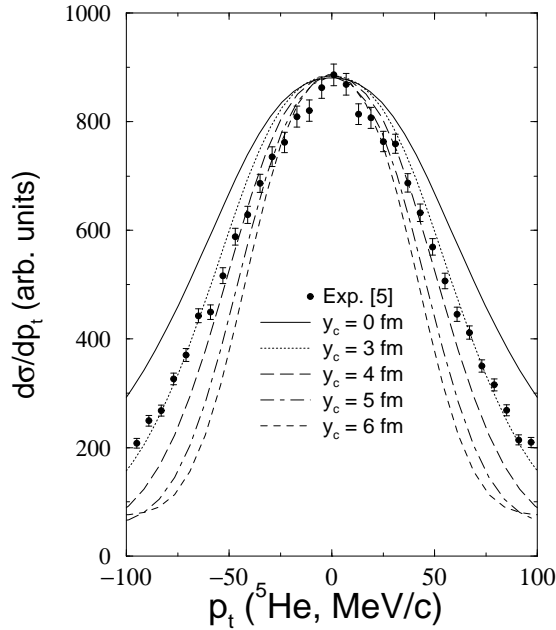,width=3in,%
bbllx=2cm,bblly=4cm,bburx=19.5cm,bbury=19.5cm,angle=-90}
\vspace{3mm}
\caption{\footnotesize Transverse $^5$He momentum distribution after
fragmentation of $^6$He on $^{12}$C at 240 MeV/u.  }
\end{figure}

The absolute value of the 2n removal cross section is sensitive to the
value of $y_c$ as seen in fig. 3a, where we show results for breakup of
$^6$He on $^{12}$C as function of beam energy for different shadowing
parameters.  The absorption processes amounts to 70-75\% of the cross
section depending on energy, consistent with the known data of neutron
scattering on $^{12}$C \cite{coo93}.  In all cases the shadowing for
$y_c=4$ fm reduces the cross section by a factor around 3.6, but the
absolute cross sections should be taken with caution due to the high
sensitivity to $y_c$. However, the numbers are encouraging and in
agreement with both experimental values \cite{tan92} and theoretical
estimates \cite{suz94,ber97}. The predicted minimum at about 200 MeV
arises from a corresponding minimum in the n-$^{12}$C cross section
\cite{coo93}.

In fig. 3b we show the value of the total 2n removal cross section,
with $y_c$ equal to 3, 4 and 5 fm relative to its value at a beam
energy of 100 MeV/u (0.430 b for $y_c=3$ fm, 0.250 b for $y_c=4$ fm
and 0.130 b for $y_c=5$ fm ).  We also show the width of the
transverse neutron momentum distribution relative to the width at 100
MeV/u (77 MeV/c, 75 MeV/c and 73 MeV/c for $y_c=$3, 4 and 5 fm,
respectively). Both these relative quantities are essentially
independent of the shadowing. In addition we note that the width of
the neutron momentum distribution is almost constant for these high
beam energies.

In fig. 4a we compare the calculated and measured transverse neutron
momentum distribution after $^6$He fragmentation on carbon. 
\begin{figure}
\psfig{file=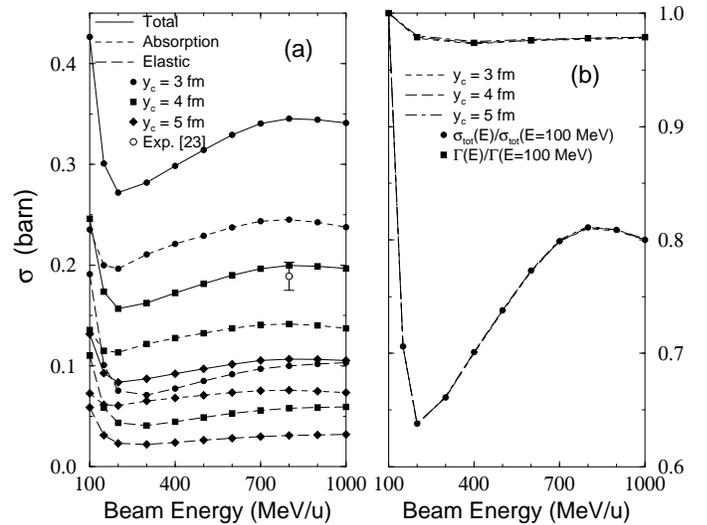,width=3.3in,%
bbllx=2cm,bblly=4cm,bburx=19.5cm,bbury=24.5cm,angle=-90}
\vspace{3mm}
\caption{\footnotesize (a) Calculated 2n removal cross section after
$^6$He fragmentation on $^{12}$C as a function of the beam energy. (b)
The same reaction as (a). The circles and squares respectively indicate
the ratio between the total 2n removal cross section and the width
of the transverse neutron momentum distribution as a function of the
beam energy.}
\end{figure}
We also show the results of calculations with and without final
state interaction based on the sudden approximation as described in
\cite{gar96,gar97}. These results resemble those without shadowing.
This fact supports the validity of the sudden approximation with inclusion
of final state interaction as a first approximation to the description
of these reactions. However, even if this approximation is rather good,
the agreement with the experiment is improved when shadowing with $y_c=4$
fm is used.
\begin{figure}
\psfig{file=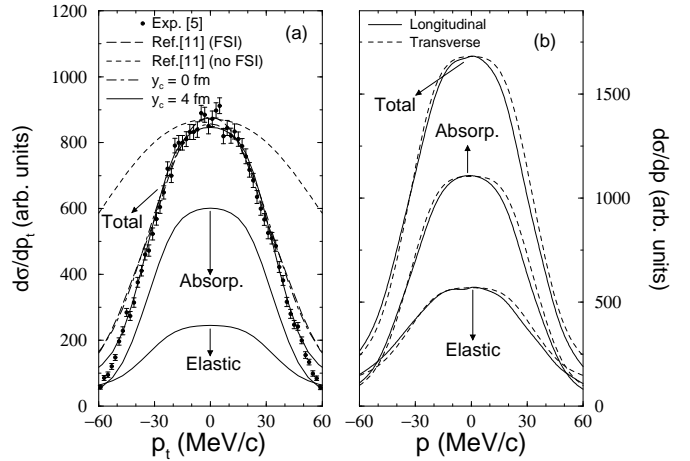,width=3.3in,%
bbllx=2.5cm,bblly=3cm,bburx=19.5cm,bbury=26.5cm,angle=-90}
\vspace{3mm}
\caption{\footnotesize (a) The transverse neutron momentum distribution
after $^6$He fragmentation on $^{12}$C at 240 MeV/u.  The results of
the calculations described in ref.[11] with (long dashed curve) and
without (short dashed curve) inclusion of final state interactions.
(b) Computed longitudinal and transverse neutron momentum distribution
for the same case as (a).}
\end{figure}

In fig. 4b we compare the calculated
longitudinal and transverse neutron momentum distributions. They are
very similar, but a small asymmetry, due to the energy dependence of
the n-$^{12}$C optical potential, is present in the longitudinal
distribution.

{\it Conclusions.} A practical method treating one particle-target
interaction at a time is formulated for spatially extended three-body
systems colliding with a light target. Absolute values of differential
cross sections including their dependence on beam energy and target
structure can then be computed. The longitudinal and transverse
momentum distributions in breakup reactions are now
distinguishable. Both diffractive and absorptive processes are
treated.. The method is subsequently applied to breakup reactions of
the halo nucleus $^6$He (n+n+$\alpha$) on a $^{12}$C target. The final
state interaction and the shadowing are crucial for the neutron
momentum distributions and the absolute cross sections respectively.
Remarkable agreement is achieved with the observed quantities for
absolute as well as for differential cross sections. If not a correct
description we provide at least a very good parametrization.

{\bf Acknowledgments.} We thank J.M. Ud\'{\i}as for providing the code
for the nucleon--nucleus potential, L.V. Chulkov and collaborators for
giving us access to the latest experimental data and K. Riisager for
continuous discussions.

\end{document}